\documentclass[journal=nalefd,manuscript=letter]{achemso}
\usepackage[version=3]{mhchem}
\usepackage{amsmath}
\usepackage{accents}

\author{Marta Castro-Lopez}
\altaffiliation{Current address: Physics Department, King's College London, London, UK}
\affiliation[ICFO - Institut de Ciencies Fotoniques, Mediterranean
Technology Park, Castelldefels (Barcelona), Spain] {ICFO -
Institut de Ciencies Fotoniques}

\author{Daan Brinks}
\altaffiliation{Current address: Faculty of Chemistry and Chemical Biology, Harvard University, Boston, USA}
\affiliation[ICFO - Institut de Ciencies Fotoniques, Mediterranean
Technology Park, 08860 Castelldefels (Barcelona), Spain] {ICFO -
Institut de Ciencies Fotoniques}

\author{\\Niek F. van Hulst}
\affiliation[ICFO -Institut de Ciencies Fotoniques, Mediterranean
Technology Park, Castelldefels (Barcelona), Spain]{ICFO -
Institut de Ciencies Fotoniques}

\alsoaffiliation[ICREA - Instituci\'o Catalana de Recerca i
Estudis Avan\c cats (ICREA), Barcelona, Spain ]{\\ICREA -
Instituci\'o Catalana de Recerca i Estudis Avan\c cats}

\email{Niek.vanHulst@ICFO.es}

\title[Non-Reciprocal Optical Antennas]
{Non-Reciprocal Optical Antennas}

\begin{document}

\begin{abstract}
Plasmonics aims to interface photonics and electronics.  Finding optical, near-field analogues of much used electro-technical components is crucial to the success of such a platform. Here we present the plasmonic analogue of a non-reciprocal antenna. For non-reciprocality in a plasmonic context, the optical excitation and emission resonances of the antenna need to be an orthogonal set.  We show that nonlinear excitation of metal nanoantennas creates a sufficient shift between excitation and emission wavelengths that they can be interpreted as decoupled, allowing for independent tuning of excitation and emission properties along different spatial dimensions. This leads, for given excitation wavelength and polarization, to independent optimization of emission intensity, frequency spectrum, polarization and angular spectrum. Non-reciprocal optical antennas of both gold and aluminum are characterized and shown to be useful as e.g. nonlinear signal transducers or nanoscale sources of widely tunable light.
\end{abstract}


\subsubsection{Introduction}

Plasmonics seeks to translate concepts of microwave and radio-frequency electronics to optical wavelengths with the aim of creating faster, smaller and more energy efficient optical-electronic interfaces. Plasmonic antennas are a highly successful exponent of this push and have been used for, among other things, enhanced or nanoscale photo detection, chemical sensing, spectroscopy, and pulse shaping \cite{Novotny2011, Brinks2013}. Like common antennas (i.e. radio- and microwave antennas), plasmonic antennas (a.k.a. nanoantennas or optical antennas) convert propagating electromagnetic radiation into electrical signals and vice versa \cite{Muhlschlegel2005b, Bharadwaj2009, Balanis2005}. In nanoantennas, the electrical signal elicited by a propagating optical wave is a plasmon, a coherent oscillation of surface electrons\cite{Ritchie1957}. Optically, this manifests as an evanescent wave in the near field of the plasmonic antenna. The most sensible definition for an optical antenna is therefore a device that converts a propagating wave into a signal in the near field, and vice versa, converts a plasmon oscillation into a propagating wave, with resonant behavior similar to that of common antennas but determined by plasmon resonances \cite{Muhlschlegel2005b, Novotny2007}.

Most types of antennas work on the principle of reciprocity: receiving a  propagating wave elicits an electrical signal from the antenna and driving the antenna with that electrical signal leads to emission of the same propagating wave. For efficient conversion of radiation into electrical signals and vice versa, it would be enough to tune the antenna resonance to this single resonant wavelength. However, the presence of active elements in the antenna system, such as amplifiers, lossy components or nonlinear elements, can breakdown this reciprocity \cite{Skolnik1990, Balanis2005}. Reception of a propagating wave then elicits an electrical signal, but driving the antenna with this signal will not lead to emission of the same propagating wave. Common non-reciprocal antennas are widely being used in, for instance, 3D radar systems where the emission needs to be as broad (spectrally and spatially) as possible while the reception must be very narrow and directed to get full knowledge of the position, shape and even composition of the target \cite{Heimiller1983,Skolnik1990}.

Non-reciprocity means that effectively, a shift has been introduced between the excitation and emission resonances of the antenna which leads to a different behavior of the system when being used as a receiving or as an emitting antenna. As a consequence, to fully characterize the response of a non-reciprocal antenna, it is necessary to measure its spectral, angular and polarization response \cite{Volakis2007} in both reception and emission working modes \cite{Visser1990}.  In Nanophotonics, the possibility of addressing emission and reception properties of antennas independently makes non-reciprocal antennas very promising for a large range of sensing, imaging, detection and signal processing applications. Multiplexing nanophotonic signals along spectral, angular and polarization axes becomes a possibility; chemical detection using plasmonics could become more accurate since both absorption and emission shifts can be quantified in independent measurements; broadband and ultrafast detection of (disturbances on optimized) ultrafast pulses could become possible; the same antennas could be used as nanoscale sources of broadly tunable light and detectors of spectrally shifted (single molecule) fluorescence.

\subsubsection{Method}

In metallic optical nanoantennas, the shorter effective wavelength of the plasmon modes compared to the incident light results in a high spatial confinement and strong local field intensity at the surface of the metal \cite{Novotny2007}. This strong local field enhancement is particularly noticeable in nonlinear interactions due to square or cubic dependence on the excitation field strength \cite{Kauranen2012}. Metallic nanostructures are therefore often characterized by methods based on nonlinear processes such as Two-Photon Photo-Luminescence (TPPL) and Second Harmonic Generation (SHG) \cite{Boyd1986, Denk1990, Beversluis2003, Krause2004, Butet2010a}. In this characterization, the intrinsic nonlinear properties of each material play a fundamental role \cite{Castro-Lopez2011}, together with the size, shape, crystallinity and oxidation state of the antenna \cite{Pelton2008, Bryant2008, Ekinci2008, Palik1985}. The fitness of metallic nanoantennas for nonlinear interaction allows implementation of the concept of non-reciprocity in plasmonic antennas based on Two-Photon absorption: specifically, two photons from a propagating wave will be absorbed by the antenna eliciting an electrical signal in the form of a plasmon oscillation (showing optically as a near field pattern around the antenna); but having that plasmon oscillation drive the nonlinear antenna will not lead to emission of a propagating wave with the same wavelength, polarization and angular spectrum as the impinging propagating wave. Metallic nanoantennas working in the nonlinear regime can thus be considered as non-reciprocal antennas. The versatility and usability of this type of nanoantennas will be determined by the separation between excitation and emission resonances: how well they are separated, how well they can be independently tuned and which properties of the excitation/emission can be individually controlled.

In this work we perform a complete characterization of non-reciprocal optical antennas. We measure the emission from these nanoantennas to establish the degree of non-reciprocity (i.e. separability between excitation and emission) by determining the strength, polarization, directionality and spectrum of the plasmonic signal compared to the propagating two-photon laser light. We find that by tuning of the different dimensions of metallic nanoantennas, we can construct non-reciprocal optical antennas for various parameters. As a first application of non-reciprocity of antennas at optical frequencies, we show how, by addressing emission and excitation resonances independently, we are able to tune the polarization, angular pattern and spectrum of the broadband TPPL emission from aluminum and gold non-reciprocal nanoantennas.

\subsubsection{Results}

Our method relies on the assumption that first order geometrical resonances of metallic nanoantennas can be approximated as dipoles in three dimensions ($\mu_{x}$, $\mu_y$, $\mu_z$) corresponding to the length, width and height ($x$, $y$, $z$) of the antenna \cite{Bryant2008, Berini2000, Schider2003, Dorfmuller2009, Taminiau2011}. A plasmon oscillation is induced through excitation with a linearly $x$-polarized laser. This oscillation is approximated as an absorbing dipole oriented in $x$-direction. The strength of this dipole will be maximum when the antenna is in resonance with the excitation wave, leading to stronger absorption and more emission; this is determined by the size of the antenna in $x$-direction as in any resonant (Fabry-Perot) cavity \cite{Jackson1998}. Two-photon absorption followed by luminescence (TPPL) is modeled as an orthogonal summation of three emitting dipoles ($\mu_{x}$, $\mu_y$, $\mu_z$) whose strengths depend on the dimensions of the antenna in each direction. Tuning the dimensions of the antenna will change the resonance condition for each dipole; weighted addition of dipoles in different directions allows control the resonant state of the antenna and therefore its angular and frequency TPPL emission spectra.

For this study, we record real space images, emission patterns (back-focal plane images, $k$-spectra), and emission spectra from nanoantennas. All of these are polarization resolved, i.e. the emission intensity in two orthogonal polarization channels ($I_{pol_x}$ and $I_{pol_y}$) is recorded for all measurements. We define the degree of linear polarization ($DoLP$) as

\begin{equation}
DoLP = \frac{I_{pol_x}-I_{pol_y}}{I_{pol_x} + I_{pol_y}}
\end{equation}

The emission detected in each polarization channel is proportional to the square of the emission dipole moment of the resonator as $I_{pol_y} / I_{pol_x}\propto \left(\mu_y / \mu_x\right)^2$. By measuring the $DoLP$ of the TPPL of the nanoantennas we can determine the relative strength of the orthogonal dipoles ($\mu_x$, $\mu_y$) via the formula

\begin{equation}\label{eq:momDoLP}
\frac{\mu_y}{\mu_x} \propto \sqrt{\frac{I_{pol_y}}{I_{pol_x}}} =
\sqrt{\frac{1-DoLP}{1+DoLP}}
\end{equation}

\begin{figure}[t]
\centering
\includegraphics[width=1.00\textwidth]{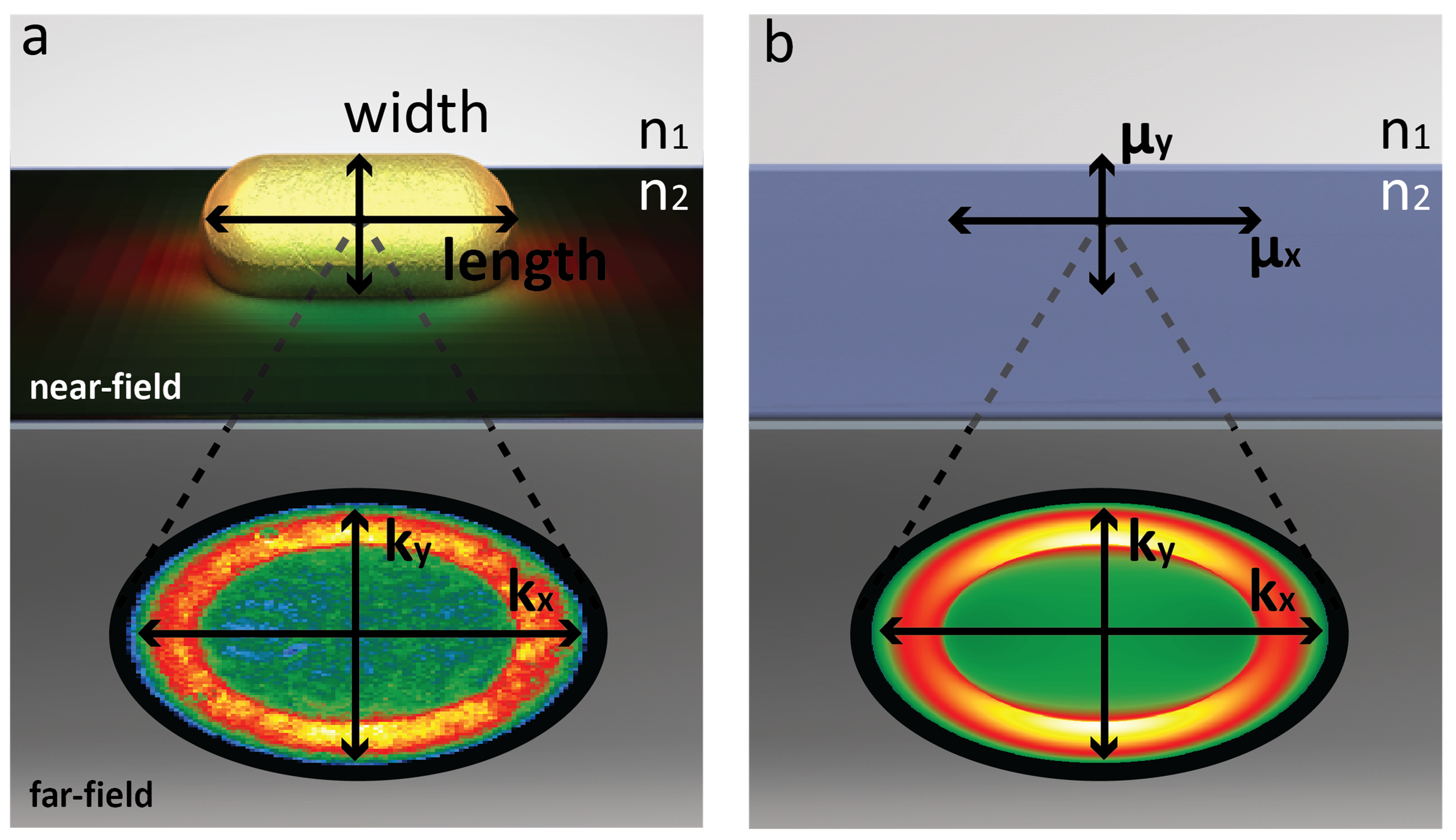}
\caption{\label{Fig-AngPattern}\textbf{Modeling geometrical
resonances of metallic nanopads.} A system of two perpendicular dipoles in the direction of the length and width of the nanorod approximate first order geometric resonances of aluminum and gold optical nanoantennas. This allows to compare experimental TPPL angular patterns with far-field theoretical predictions.}
\end{figure}

In a high NA objective, some polarization mixing will occur, i.e. the signal detected in the far field after filtering out a particular polarization will contain some deterministic contribution of orthogonal polarizations.  To calculate the relative strengths of the emission dipoles (Equation 2) we need to compensate for the cross-polarization terms (see supplementary information). Figure 1 shows a schematic of the system and method under consideration: a plasmonic antenna is modeled as a set of orthogonal dipoles on a glass-air interface, inspected from the glass side via an objective. The weighted addition of the dipoles leads to an angular spectrum that can be compared with the measured back-focal plane images. After proper compensation for cross-polarization components, measured values of $\mu_x/\mu_y$ are used to simulate the angular spectrum of the antenna under investigation.

The nanoantennas are fabricated on top of a thin film of ITO (10 nm) by electron beam lithography using PMMA as positive photoresist. After development, 40 nm of either gold or aluminum are thermally evaporated at 1.5 $\mathring{A}/s$, followed by lift-off with dichloromethane. The width of the nanoantennas is well-defined down to 50 nm. We choose the antennas to have a constant height of $z$=40 nm and let the antenna length be determined by the excitation resonance condition to drive the plasmon oscillation with highest efficiency ($x$=110 nm for gold antennas and $x$=160 nm for aluminum antennas). To investigate the non-reciprocity of antennas of different dimensions, we vary the width of the antennas in steps of 10 nm between 50 and 300 nm. The antennas are excited with $x$-polarized femtosecond pulses at $\lambda_{exc}$=800 nm. The TPPL band between 440 and 690 nm is collected using an oil immersion objective (Zeiss, plan-Apo 100x/1.46 NA) and redirected towards three different detection paths which allow polarization resolved real space imaging, angular spectrum imaging and frequency spectral measurements (see supporting information). In this manner, similar to how non-reciprocity is measured in radio/micro wave antennas, a complete characterization of the non-reciprocity of gold and aluminum nanoantenna systems is achieved.

Figure 2 (top) shows the real space confocal image of a series of aluminum nanoantennas of 160 nm length with increasing widths from 50 nm (left) to 300 nm (right) in steps of 10 nm. The color codifies the $DoLP$ from -1 to 1 (green to red). As an example, an aluminum nanoantenna of dimensions 160$\times$50$\times$40 nm ($x$, $y$, $z$) being excited at $\lambda_{exc}$=800 nm presents a $DoLP$=0.8. This indicates a pure $x$-oriented dipole (see supporting information). Figure 2(a) show the associated angular spectrum, which can be decomposed into $x$-polarized (Figure 2(b)) and $y$-polarized (Figure 2(c)) components by placing a polarizing beam-splitter in front of the camera and recording the back-focal plane image for two orthogonal polarizations, parallel with the excitation dipole ($x$) and perpendicular with the excitation dipole ($y$). Note that Figure 2(a) is reconstructed from the $x$- and $y$-components by addition after weighing by a correction factor that compensates for non-ideal transmission through the polarizing beam splitter. Each of them is measured by integrating photons for 30 seconds.

TPPL photons in the 440-690 nm band couple to the modes of the antenna and are emitted following the characteristic angular pattern of the mode. Only light emitted towards angles $\leq ($NA$/n_{glass})$ is collected, causing the sharp circular cut-off in the back focal plane images. The far-field angular spectrum of the $x$-polarization component (Figure 2(b)) is a clear signature of the emission of a dipole oriented in the $x$-direction. The $y$-polarization component (Figure 2(c)) corresponds to the $y$-projection (cross-polarization) of an $x$-oriented dipole in a high NA objective (glass interface). Since the system is not in resonance with the 440-690 nm TPPL emission, the polarization and angular spectrum take on the characteristics of the excitation dipole driven by the linearly polarized excitation laser. We can compare this with theory by simulating two perpendicular dipoles with relative strength given by Equation 2. After compensation for polarization mixing, the values fed into the simulations are $\mu_x$=0.97 and $\mu_y$=0.15. The simulated far-field angular spectra are shown in Figures 2(d-f) and form a near perfect match with the measured angular spectrum. The implication of this is that for these dimensions ($x$, $y$, $z$ = 160, 50, 40 nm) the aluminum antenna behaves in many respects as a reciprocal antenna for polarization and angular spectrum but not frequency: a propagating wave incites an electrical signal which maintains the angular spectra and the polarization of the propagating wave but blue-shifts its wavelength.

\begin{figure}[t]
\centering
\includegraphics[width=0.99\textwidth]{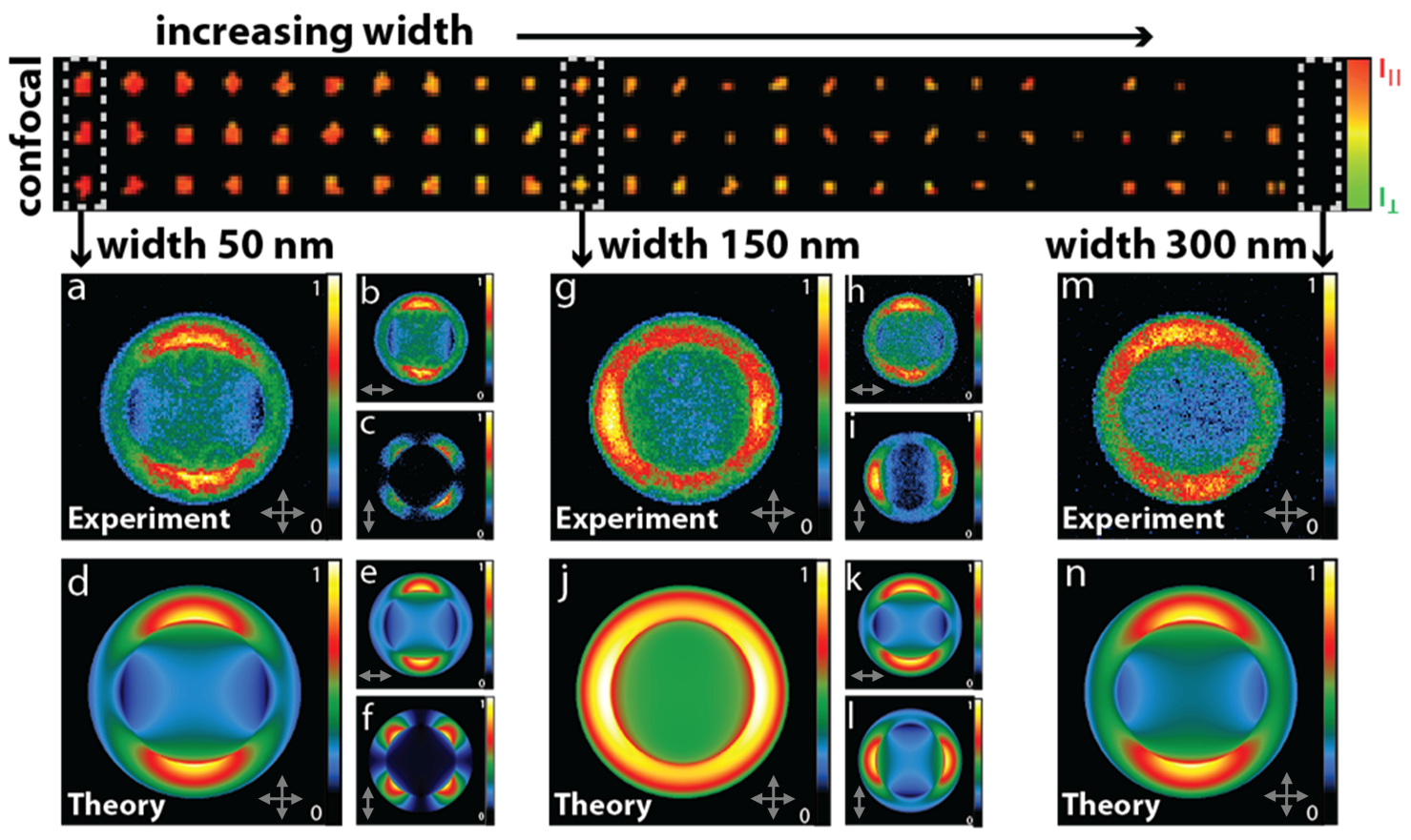}
\caption{\label{Fig-AngPattAl}\textbf{Control of angular pattern
and polarization of the TPPL emission of aluminum nanopads.} Top
panel: confocal image of the TPPL of aluminum nanopads of similar
length $\sim$160 nm and widths increasing from 50 nm (left) to 300
nm (right) in steps of 10 nm. Color codifies polarization of the
TPPL signal. Bottom panel: measured (top row) and simulated
(bottom row) angular patterns of aluminum nanopads of widths 50 nm
(left), 150 nm (middle) and 300 nm (right). (a), (d), (g), (j),
(m) and (n) show the full angular patterns. (b), (e), (h) and (k)
show the $x$-projection of the polarization. (c), (f), (i) and (l)
show the $y$-projection of the polarization. }
\end{figure}

To increase the non-reciprocity of the system (i.e. separability between propagating wave and electrical signal) we vary the width of the nanoantenna. FDTD simulations show that the resonance length for an aluminum particle at $\lambda$=600 nm is 135 nm. Increasing the width of the antenna we observe a drop in $DoLP$ from $DoLP$=0.8 @ 50 nm width to a minimum of $DoLP$=-0.1 @ 160 nm width (Figure 2, top). Figures 2(g-l) show the measured (g-i) and simulated (j-l) angular spectra associated with an antenna of 150 nm width. The $x$-projection (Figure 2(h)) stays very similar to the previous case (Figure 2(b)), indicating an $x$-oriented dipole contributing to the emission. On the other hand, the $y$-projection (Figure 2(i)) now also shows a clear dipolar pattern. Weighted addition of both projections, to create the full angular pattern (Figure 2(g)), shows that the angular spectrum has developed into a more homogeneous ring with lobes of maximum emission rotated 90$\mathring{}$ compared to the previous case. Using the $DoLP$ obtained from Figure 2 (top) it is determined that $\mu_y\approx$1.1$\mu_x$. After correction for cross-polarization terms, $\mu_x$=0.82 and $\mu_y$=0.95 are used to simulate the angular spectra in Figures 2(j-l).

Increasing the width of the antenna even further, past the dimensions the antenna's width is resonant with the emission spectrum, returns the polarization and angular spectrum to that of a single $x$-oriented dipole (Figures 2(m,n)). Such a nanopad presents a $DoLP$=0.52 @ 300 nm width, which gives a relative strength between $x$- and $y$-oriented dipoles as $\mu_y\approx$0.56$\mu_x$.

While aluminum antennas show some non-reciprocal behavior with interesting broadband dipolar properties and are attractive because of the low price and abundance of the material, gold is a more familiar material for plasmonics. Previous polarization analysis of gold antennas shows strong dependence of emission polarization on antenna dimensions \cite{Castro-Lopez2011}. It is therefore likely that gold nanoantennas will exhibit desired non-reciprocal behavior.

Figure 3 (middle) shows the real space confocal image of a series of gold nanoantennas of 110 nm length with increasing widths from 50 nm (left) to 300 nm (right) in steps of 10 nm. The color codifies the $DoLP$ from -1 to 1 (green to red). It is immediately clear that the polarization response to changes in the antenna width is more pronounced than for aluminum. This is corroborated in the angular emission spectra. Figure 3(a) shows the angular emission spectrum of a gold nanoantenna of ($x$, $y$, $z$) = (110, 50, 40) nm excited at $\lambda_{exc}$=800 nm.

The circular emission pattern cannot be explained by a multipole interaction; instead, the uniform distribution of intensity in both polarization and emission angles points in the direction of multiple orthogonal dipoles contributing to the emission. Decomposition of the emission pattern in $x$- (Figure 3(b)) and $y$-polarization (Figure 3(c)) shows identical, but 90$\mathring{}$ rotated angular spectra. These can be accurately fitted (Figures 3(d-f)) by modeling two orthogonal dipoles oriented along $x$ and $y$. Note that the subtle difference between pure $x$-dipolar emission in the $x$-polarization channel and the signal recorded here can completely fitted by addition of the $x$-polarized component of the emission of a $y$-oriented dipole; no multipole expansion is necessary.

\begin{figure}[H] \centering
\includegraphics[width=0.99\textwidth]{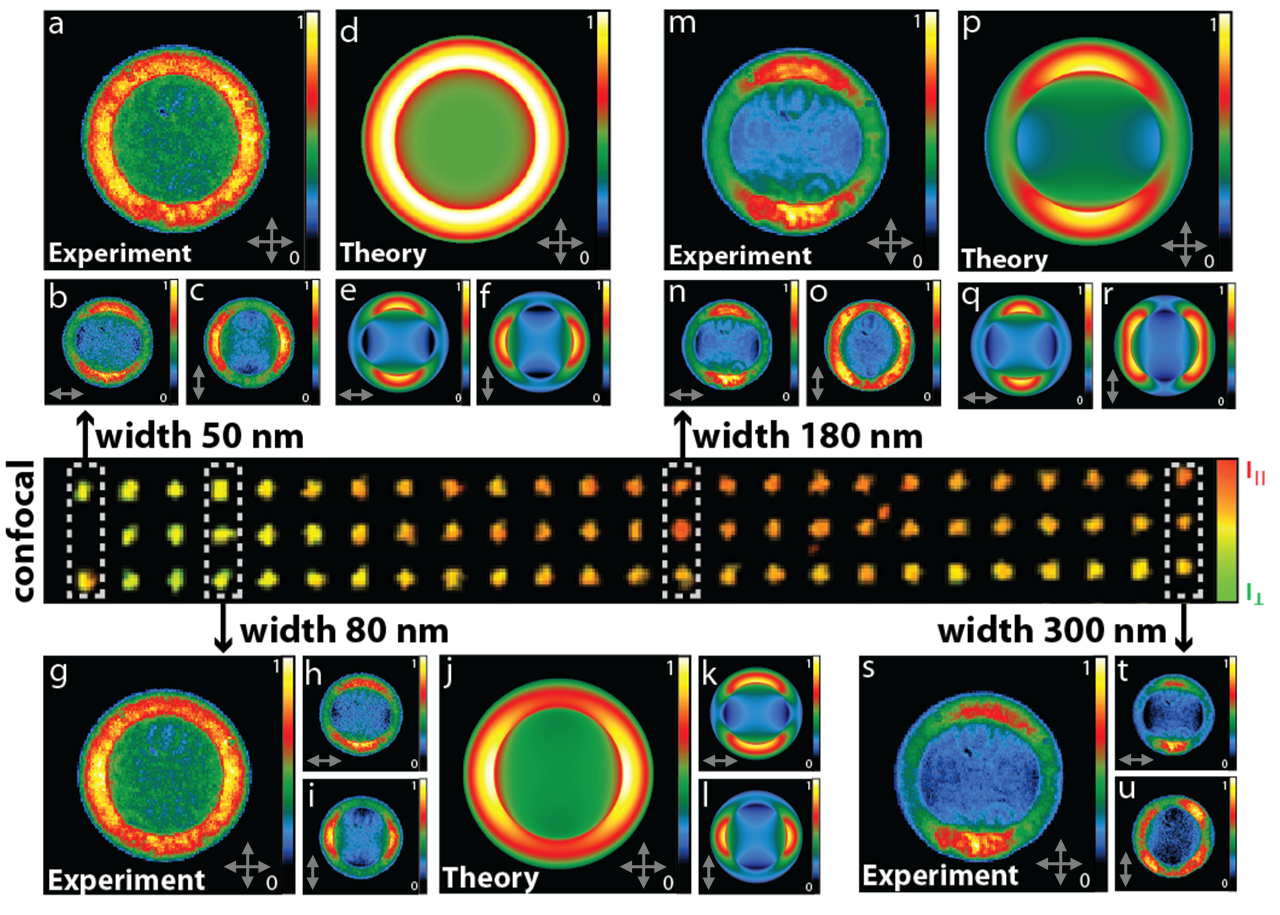}
\caption{\label{Fig-AngPattAu}\textbf{Control of the angular
pattern and polarization of the TPPL emission of gold nanopads.}
Middle panel: confocal image of the TPPL of gold nanopads of
similar length $\sim$110 nm and widths increasing from 50 nm
(left) to 300 nm (right) in steps of 10 nm. Color codifies
polarization of the TPPL signal. Top and bottom panels: measured and simulated angular patterns of gold
nanopads of widths 50 nm (top-left), 80 nm (top-right), 180 nm
(bottom-left) and 300 nm (bottom-right). (a), (d), (g), (j), (m),
(p) and (s) show the full angular patterns. (b), (e), (h),
(k), (n), (q) and (t) show the $x$-projection of the
polarization. (c), (f), (i), (l), (o), (r) and (u) show the
$y$-projection of the polarization. Starting from an homogeneous
ring in the case of a width of 50 nm (a), the angular pattern
becomes that of a dipole in the $y$-direction when the
width is 80 nm (g) and rotates 90$\mathring{}$ to the $x$-direction when the width increases further (m).}
\end{figure}

The appearance of a strong second dipole in the emission ($\mu_y$) opens the way for further control of the non-reciprocal behavior of gold antennas. The strength of this second, orthogonal dipole can be tuned by varying the width of the antenna, as demonstrated in Figures 3(g-l), where the $y$-oriented dipole is dominant, and Figures 3(m-q) where the $y$-oriented dipole has almost completely been turned off. The corresponding dipole strengths are $\mu_y$ $\approx$ $\mu_x$, $\mu_y$ $\approx$ 1.22$\mu_x$ and $\mu_y$ $\approx$ 0.67$\mu_x$ for 50, 80 and 150 nm widths.

We have shown how one of the dimensions along which orthogonality (leading to non-reciprocal behavior) can be established, is directionality: proper filtering can establish detection of an emitted signal that is, in terms of angular content, the inverse of the signal received by the antenna, as demonstrated in Figures 2 and 3. Another possible dimension where orthogonality can be established is the frequency spectrum. While the TPPL excitation-emission inherent to our scheme already establishes a shift between propagating wave (800 nm) and detected signal (440-690 nm), combining the spectral and directional channels leads to a higher degree of non-reciprocality. For different applications of the antennas, these different channels could be used for multiplexing the signal, or could be combined for higher fidelity.
The analysis performed in Figures 2 and 3 establishes, for each set of antenna dimensions, a tunable relation between polarization and angular content of the emission. Recording polarization resolved frequency spectra gives us, via this polarization relation, enough information to reconstitute the ($k$, $\omega$)-spectrum of the emission of the non-reciprocal antennas.

A more in depth analysis is performed in figures 4 and 5. Figure 4(a) shows the spectra of $x$-polarized emission of gold antennas of different widths; Figure 4(b) shows the same for the $y$-polarized emission. The peak shift in the $x$-polarized emission is attributed to the cross-polarization of the $y$-oriented dipole in a high NA objective, when the increase in width shifts the antenna in and out of resonance with the intrinsic TPPL emission spectrum. Two resonant modes can clearly be observed in the $y$-polarized emission. Increasing the width beyond 230 nm leads to the appearance of a second order resonance mode around 600 nm.

\begin{figure}[t]
\centering
\includegraphics[width=0.99\textwidth]{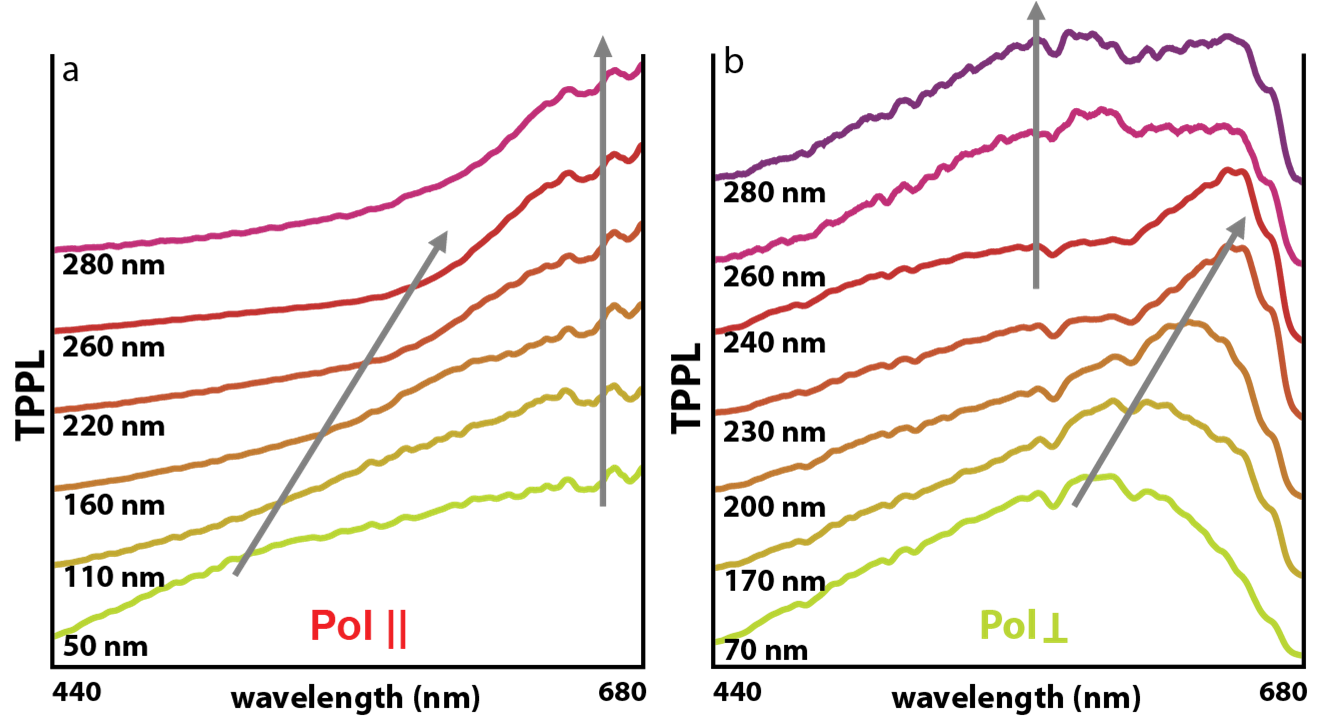}
\caption{\label{Fig-SpectrumAu}\textbf{Control of TPPL emission
spectra of gold non-reciprocal nanoantennas.} (a) Spectral evolution of the $x$-polarized TPPL emission from a nanopad with increasing width (50-280 nm). The spectral maximum is outside the detection window (right arrow), which leads to a an artificially stable spectral maximum. The diminution of the counts around 600 nm (left arrow) is attributed to the disappearance of the small contribution of the cross-polarization of a $y$-oriented dipole. (b) Spectral evolution of the $y$-polarized TPPL emission for gold nanopads of widths 70 nm, 170 nm, 200 nm, 230 nm, 240 nm, 260 nm and 280 nm. As the width increases, the peak of the spectrum (right arrow) is moving towards higher wavelengths and disappearing from the detection window. When further increasing the width (>230 nm) a second order resonance mode of the emission starts to appear around 600 nm (left arrow). }
\end{figure}

From Figure 4 it is clear that $x$- and $y$-polarized emission show different spectra, leading to the possibility of combined directional and spectral control; different parts of the angular spectrum of the emission will contain different spectral components. This concept is elaborated upon in Figure 5, where we show two examples of full multi-parameter control of non-reciprocal antennas. In particular, Figure 5(a) shows the spectra for $x$- and $y$-polarized emission for a gold nanoantenna of ($x$, $y$, $z$)=(110, 70, 40) nm. From Figures 3(b-c) it is established that those emission bands are dominated by dipolar angular spectra, oriented in $x$- and $y$-direction. This is confirmed in Figure 5(b), where the angular pattern, color coded for polarization (red represents Pol$_x$ and green Pol$_y$; yellow represents a mix of polarizations), shows how different polarizations, containing different spectral components, are emitted at different angles. The emission in $y$-direction is for instance dominated by light with wavelength around 650 nm, whereas the emission in $x$-direction is dominated by a broad spectrum between 440 and 600 nm.

This complete information about the spectral and spatial content of the TPPL emission of a non-reciprocal nanoantenna now allows us to imagine a new use for such a device: that of nanometrically, substructure-localized tunable excitation source. Figure 5(c) shows the spatial near-field distribution around the nanorod investigated in 5(a-b),obtained by FDTD simulations (color coded in the same way as figure 5b). It is obvious that different modes are located at different positions on the nanorod; excitation of the rod with 800 nm light and positioning a molecule or other nanosized absorber near the antenna would give us sub-antenna sized localization accuracy based on the relative excitation probability of the system, or the possibility to multiplex detection, i.e. enhance the detection of different molecular species with the same antenna, excited with the same 800 nm light.

 \begin{figure}[t]
\centering
\includegraphics[width=0.99\textwidth]{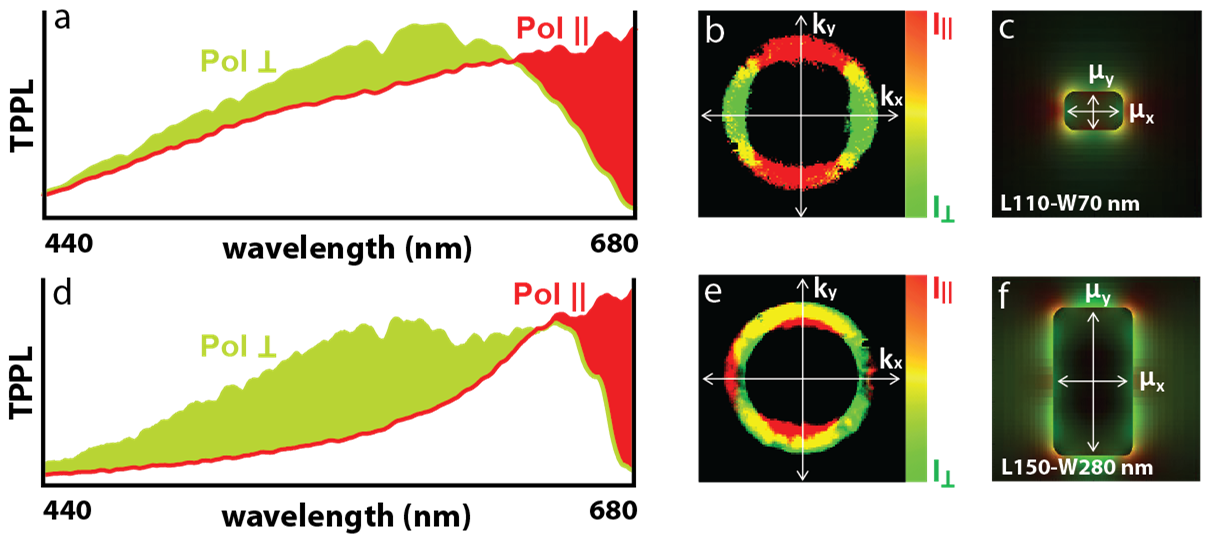}
\caption{\label{Fig-NonRecAnt}\textbf{Control of nonreciprocal antennas.} (a) TPPL spectra of
a gold nanopad resonant with the excitation (length = 110 nm) and
the emission (width = 70 nm). The $y$-projection of the
polarization (Pol$\perp$) shows a strong broad resonance peak
around 600 nm wavelength while the peak of the $x$-projection of
the polarization (Pol$\parallel$) is filtered out. (b) TPPL spectra of
a gold nanopad slightly off resonant with the excitation (length = 150 nm) and passing through the second order
resonance with the emission (width = 280 nm).}
\end{figure}

The tenability of this aspect of non-reciprocal antennas is illustrated in Figures 5(d-f), where the spectra, angular pattern and near-field distribution for a gold nanoantenna of ($x$, $y$, $z$)=(150, 280, 40) nm are shown. In this case, $x$- and $y$-polarized emission seem to be largely disentangled from each other (Figure 5(d)). Indeed, the TPPL signal at lower wavelengths is heavily dominated by $y$-polarized emission while at larger wavelengths the $x$-polarized takes the lead. The complexity of the angular pattern shown in Figure 5(e) has increased compared to previous case and shows the different angles at which lower and higher wavelengths are emitted (green and red). The spatial distribution of the near-field obtained by FDTD simulations also shows an increased complexity. Different sub-diffraction regions around the nanoantenna containing different wavelengths are routed towards different angles. These examples shows how the emission from non-reciprocal antennas can be spatially, angularly and spectrally controlled independent from the excitation.

Interestingly, our findings for gold and aluminum suggest different uses for both of them: aluminum antennas are able to achieve a much purer dipole-like emission, whereas the tunability of the emission properties of gold antennas is much larger. For aluminum antennas, it is therefore feasible to talk about a broadband dipole with a bandwidth of more than 200 nanometers, which is eminently suited for ultrafast plasmonics. On the other hand, gold antennas give better combined angular-spectral control of the emission which is better for chemical detection, positioning, or signal converters (e.g. angular conversion of wavelength conversion). 

In conclusion, we introduced the equivalent of a non-reciprocal antenna at optical frequencies. Our approach was based on nonlinear excitation of gold and aluminum plasmonic nanoantennas. The versatility of these systems was shown by tuning the angular and spectral properties of their emission modes independent of their reception mode. Broadband and tunable emission modes were demonstrated, as well as the possibility of either increasing emission or detection fidelity by combining multiple channels, or multiplexing  detection/emission possibilities along different channels. The full spatial, spectral and angular information for these antennas was shown, allowing the envisioning of improved chemical detection, localization and signal conversion applications, which, depending on the material, could be used in ultrafast or high-spatial resolution applications.

\begin{acknowledgement}
The authors thank T. H. Taminiau for the initial source code used
for the simulations and R. Sapienza for the 3D rendering. D.B. acknowledges support by a Rubicon grant of the Netherlands Organization for Scientific Research (NWO) and a  Howard Hughes Medical Institute (HHMI) fellowship.
\end{acknowledgement}


 \providecommand*\mcitethebibliography{\thebibliography}
\csname @ifundefined\endcsname{endmcitethebibliography}
  {\let\endmcitethebibliography\endthebibliography}{}


\newpage

\renewcommand{\thefigure}{S.\arabic{figure}}
\setcounter{figure}{0}  

\section{Supplementary Information}

\subsubsection{Modeling non-reciprocal antennas}
In any resonant (Fabry-Perot) cavity, the position of the resonance modes is determined by the size of the cavity ($S_{res}$)
\begin{equation}
S_{res}(m)= \frac{m\pi}{k_{eff}}    ; m = 1, 2, 3 ...
\end{equation}
Being $k_{eff}$ the wavevector of the mode and $m$ the number of the order \cite{Jackson1998}. Therefore, in the approximation of modeling metallic nanopads as lossy optical cavities \cite{Berini2000, Schider2003, Bryant2008, Dorfmuller2009, Taminiau2011}, the actual size of the nanorod on each direction (length, width and height) will determine its resonance state, together with excitation/emission wavelengths. In this case, $k_{eff}$ will be a complicated function depending on the dispersive properties of the material of the nanostructure ($\varepsilon(\lambda)$) \cite{Novotny2007}. For the dimensions and materials of the structures under study, the $S_{res}(m)$ of the nanopads under study will correspond to the first dipolar mode ($m = 1$).

\begin{figure}[t]
\centering
\includegraphics[width=1.00\textwidth]{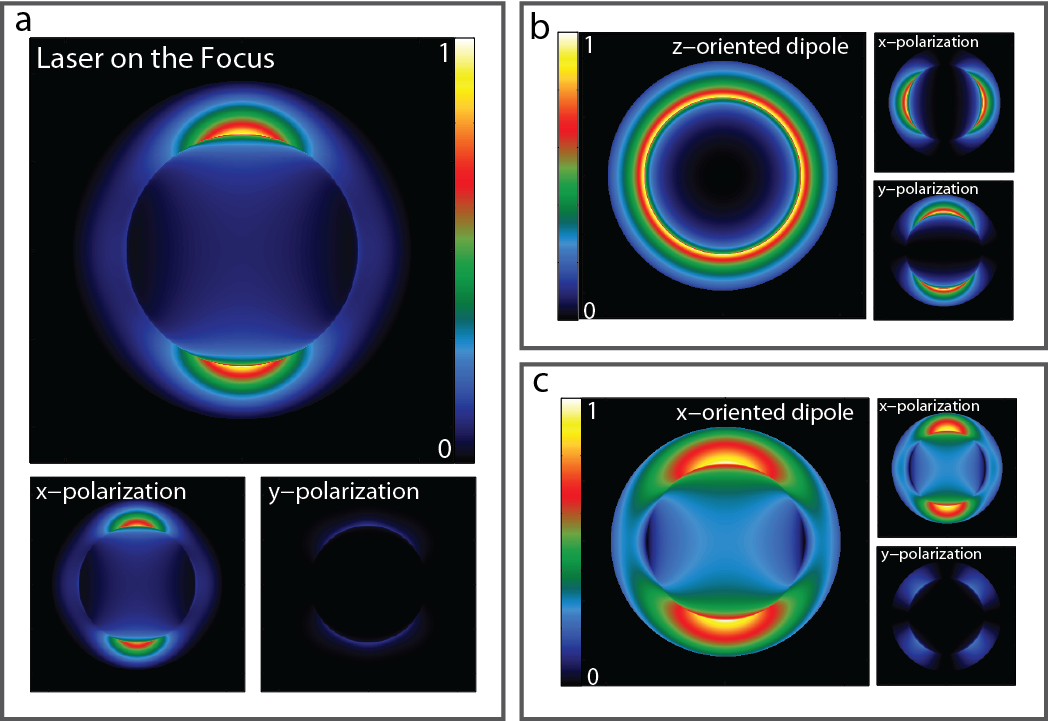}
\caption{\label{Fig-Consider}\textbf{Considerations for angular pattern measurements.}
(a) Simulation of the angular pattern of a dipole representing the
excitation seen at the focus of the objective when excited by
two-photon absorption with a linear $x$-polarized laser. Most
excitation occurs in the $x$-direction as the $y$-projection of the
polarization is very low compared to the $x$-projection. (b)
Simulation of the angular pattern of a $z$-oriented dipole. Photons
are equally distributed between $x$- and $y$-projections. (c)
Simulation of the angular pattern of a $x$-oriented dipole. Most of
the photons are emitted with $x$-polarization although the
y-projection has also a representative pattern. }
\end{figure}

While we have in principle enough information to separate contributions from $\mu_x$, $\mu_y$ and $\mu_z$, only in-plane dipoles ($x-$ or $y$-oriented dipoles) will be considered. Although in a high NA objective it is possible to have vertical components of the excitation field \cite{Erikson1994} 
($E_z \propto
\frac{\omega_\circ}{\lambda}E_x$, being $\omega_\circ$ =
2$\frac{\lambda}{\pi NA}$ the beam waist), our
excitation is quadratic (two-photon absorption) thus the relative strength of any absorbing z-dipoles will be a factor of
$\left(\omega_\circ / \lambda\right)^{4}$ smaller than that of $x$-dipole components (the relative intensity exciting $y$ dipoles will
be a factor of $\left(\omega_\circ / \lambda\right)^{8}$). Thus, $\mu_z$ is negligible in the excitation process. (Figure~\ref{Fig-Consider} (a)). At the same time, the height of all the structures is kept constant for all the experiments presented here, so the presence of any resonance in the direction of $z$ will contribute equally to both $x$ and $y$ polarizations (Figure~\ref{Fig-Consider} (b)).

\subsubsection{Experimental methods}
The setup used for the characterization of non-reciprocal optical nanoantennas presents three different detection branches which allow to measure polarization, angular pattern and spectrum of the emitted TPPL respectively. On the first path, a polarization beam cube in front of a
pair of APDs (Single Photon Avalanche Diode, SPAD) gives a
polarization resolved real space image. Each point/pixel (100 nm$^{2}$)
in the real space image is integrated during 1.8 ms. On the second path, a conoscope performs an image of the back focal plane of the objective, where
the Fourier plane is in focus i.e each point of the plane
represents a single emitted angle. A rotating broadband linear polarizer
integrated in the conoscope before the EMCCD camera (Hamamatsu
ImagEm C9100-13) gives a polarization resolved momentum space
image. The
momentum space images are given by traces of 30 seconds for each
polarization. And finally, in the third path the TPPL signal can be directed towards a spectrometer (Andor Shamrock
303, Newton EMCCD). Again a broadband linear polarizer in front of
it allows to perform polarization resolved spectral images. Each spectrum is given by
integrating the arriving photons during traces of 10 or 30 seconds
depending on the resonance strength/state of the particular mode
being measured.

\subsubsection{Angular pattern simulations}
The angular pattern simulations are done calculating the far-field of a dipole with momentum $\mu$ = ($\mu_x$,$\mu_y$,$\mu_z$). This dipole is placed in a medium of index $n_1$ (in the experiments $n_{air}$ = 1)  and it is located 10 nm above the origin (($x$,$y$)=(0,0)) of an interface defined by the plane $z$ = 0. The medium below the interface ($z$ $\le$ 0) is defined to have a different refractive index $n_2$ (in the experiments $n_{glass}$ = 1.52). Since the measurements done in this chapter are reflexion measurements, this medium is taken as the direction of observation. The wavevector at both sides of the interface is
\begin{equation}
k_i = \frac{2\pi}{\lambda}n_i ; i\in{1,2}
\end{equation}

Since the angular spectrum will depend on the projection of $k$ on the $z$ axis ($k_z$) towards both mediums but seen from medium 2, we define
\begin{equation}
k_{z2} = k_2cos\theta
\end{equation}
and
\begin{equation}
k_{z1} = \sqrt{k_1^2-k_2^2sin^2\theta}
\end{equation}

The transmission Fresnel coefficients for p- and s-polarized light, respectively are then:
\begin{equation}
t_p = \frac{2\epsilon_1k_{z2}}{\epsilon_1k_{z2} + \epsilon_2k_{z1}}\frac{n_2}{n_1}
\end{equation}
and
\begin{equation}
t_s = \frac{2k_{z2}}{k_{z2}+k_{z1}}
\end{equation}

The field emitted/transmitted to this medium is weighted by these coefficients. In the case of a vertical dipole ($\mu$ = (0, 0, $\mu_z$)), the transmitted potential only contains p-polarized light of the form:
\begin{equation}
\phi_V = -\frac{n_2^2}{n_1^2}t_pe^{ik_{z1}z_0}
\end{equation}

For the case of an horizontal dipole ($\mu$ = ($\mu_x$,$\mu_y$,0)) there are p and s contributions respectively of the form:
\begin{equation}
\phi_{H_p} = - \frac{n_2}{n_1}t_pe^{ik_{z1}z_0}
\end{equation}
and
\begin{equation}
\phi_{H_s} = - \frac{n_2}{n_1}t_se^{ik_{z1}z_0}
\end{equation}

The minus sign in $\phi_V$ and $\phi_{H_s}$ is due to the direction of observation -$z$ given by the substitution:
\begin{equation}
\frac{k_{z2}}{k_2} = -cos\theta
\end{equation}
while
\begin{equation}
\frac{k_{z1}}{k_1} = cos\theta
\end{equation}

Ignoring the common therms $\frac{k_1^2}{4\pi\epsilon_0\epsilon_1}\frac{e^{ik_2}}{r}\frac{-n_2}{n_1}$ between $E_\theta$ and $E_\phi$ it is possible to write:
\begin{equation}\label{Etheta}
E_\theta = -(\mu_xcos\phi + \mu_ysin\phi)\frac{k_{z1}}{k_1}\phi_{H_p} - \mu_zsin\theta\frac{k_2}{k_1}\phi_V
\end{equation}
and
\begin{equation}\label{Ephi}
E_\phi = (-\mu_xsin\phi + \mu_ycos\phi) \phi_{H_s}
\end{equation}

Coming back to cartesians:
\begin{equation}
E_x = E_\theta cos\theta - E_\phi sin\theta
\end{equation}
and
\begin{equation}
E_y =  E_\theta sin\theta + E_\phi cos\theta
\end{equation}

It is possible now to write the total power emitted into this lower medium as
\begin{equation}
P_{TOT} = \frac{\left|E_\theta\right|^2+\left|E_\phi\right|^2}{cos\theta} = \frac{\left|E_x\right|^2+\left|E_y\right|^2}{cos\theta}
\end{equation}

In order to distinguish between $x$ and $y$ polarizations it is possible to write:
\begin{equation}
P_{x,y} = \frac{\left|E_{x,y}\right|^2}{cos\theta}
\end{equation}

Finally, since the measurements are done using a high NA objective (NA = 1.46), there is a limitation on the angles $\theta$ that are possible to detect. Therefore, the power emitted towards certain angles will be lost, those angles are:
\begin{equation}
\theta \geq arcsin\frac{NA}{n_2} \approx 73.8\mathring{}
\end{equation}

It is now straightforward to calculate the radiation pattern of a particular set of perpendicular dipoles just by setting specific values of $\mu_x$, $\mu_y$ and $\mu_z$ in equations (14) and (15) and calculating the power emitted via $P_{x,y}$. Note that only in plane dipoles ($\mu_x$ or $\mu_y$ dipoles) will be considered (Supplementary Information, Modeling non-reciprocal antennas).

\subsubsection{Cross-polarization correction}
The relative field between channels $x$ and $y$ is related to the measured intensity and the DoLP as
\begin{equation}
\frac{E_y}{E_x} \propto \sqrt{\frac{I_y}{I_x}} = \sqrt{\frac{1-DoLP}{1+DoLP}}
\end{equation}

Since the far field of a pure $x$-oriented dipole has also components on the $y$-polarization (Figure~\ref{Fig-Consider} (c)), in order to obtain the values of the dipoles from far field polarization measurements we need to compensate for these cross-polarization terms. Therefore the values used for the simulations ($E_{sx}$ and $E_{sy}$) are
\begin{equation}\label{eq1}
E_{sx} + (\frac{w_0}{\lambda})^2 E_{sy} = E_x
\end{equation}
\begin{equation}\label{eq2}
E_{sy} + (\frac{w_0}{\lambda})^2 E_{sx} = E_y
\end{equation}

Being $w_0$ the beam waist. Since we are interested in the relative strength between channels, we fix $E_x = 1$ so $E_y=\sqrt{(1-DoLP)/(1+DoLP)}$. As a result, (22) and (23) become a system of two equations with two unknowns, therefore by substitution we get
\begin{equation}
E_{sx} = \frac{1-((\frac{w_0}{\lambda})^2\sqrt{\frac{1-DoLP}{1+DoLP}})}{1-(\frac{w_0}{\lambda})^4}
\end{equation}
And
\begin{equation}
E_{sy} = \sqrt{\frac{1-DoLP}{1+DoLP}} - (\frac{w_0}{\lambda})^2 E_{sx}
\end{equation}

Finally, from the relation $E_{sx} \propto \mu_{sx}$ we extract the relative strength between $\mu_{sx}$ and $\mu_{sy}$.

\subsubsection{Comparing angular patterns}
In the comparison between simulated and measured angular patterns, two facts have to be taken into account. First, the DoLP used to calculate the relative strength between $x-$ and $y-$oriented dipoles by
definition can reach values between 1 and -1 (completely
longitudinal or transversal polarization respectively), however,
the presence of a glass interface close to the dipoles make it not
possible to have pure s or p-polarized light. The real range of
values of DoLP is between 0.8 and -0.8 \cite{Erikson1994} . 
The theory
behind the simulations has, by definition, considered this aspect. The second fact to take into account is the presence of the 50/50
beam splitter used to separate the TPPL signal. The s- and
p-polarization transmission of such beam splitters is different. We will need to introduce a correction factor on the
measured angular patterns. Since we use a linear polarizer to
measure, this correction factor will only be taken into account
when adding angular patterns of perpendicular polarizations to
obtain the total angular pattern.

\end{document}